\documentclass[letterpaper, 10 pt, conference]{ieeeconf}  

\IEEEoverridecommandlockouts                              

\usepackage{booktabs} 
\usepackage{amsmath}
\usepackage{multirow}
\usepackage{amssymb}
\usepackage{graphicx}
\usepackage{wrapfig}
\usepackage{subfigure}
\usepackage{url}
\usepackage{cite}
\usepackage{textcomp}
\usepackage{verbatim}
\usepackage{gensymb}
\usepackage{array}
\usepackage{amssymb}
\usepackage{multirow}
\usepackage{footnote}
\usepackage{empheq}

\newtheorem{Theorem}{Theorem}

\title{Stochastic Virtual Battery Modeling of Uncertain Electrical Loads using Variational Autoencoder*
}

\author{Indrasis Chakraborty\textsuperscript{1}, Sai Pushpak Nandanoori\textsuperscript{2}, Soumya Kundu\textsuperscript{2}, and Karanjit Kalsi\textsuperscript{2}
\thanks{*This work was supported by the Grid Modernization Initiative of the U.S. Department of Energy under the contract DE-AC05-76RL01830.}
\thanks{1 Indrasis Chakraborty is with the Center for Applied Scientific Computing, Lawrence Livermore National Laboratory, Livermore, CA 94550, USA. E-mail contact: {\texttt\small chakraborty3@llnl.gov}, and was with Pacific Northwest National Laboratory during the execution of the work.}%
\thanks{2 Sai Pushpak Nandanoori, Soumya Kundu, and Karanjit Kalsi are with the Optimization and Control Group, Pacific Northwest National Laboratory, Richland, WA 99352, USA. E-mail contacts: {\texttt\small saipushpak.n@pnnl.gov, soumya.kundu@pnnl.gov, karanjit.kalsi@pnnl.gov}}
}

\begin{document}

\maketitle
\thispagestyle{empty}
\pagestyle{empty}

\begin{abstract}
Effective utilization of flexible loads for grid services, while satisfying end-user preferences and constraints, requires an accurate estimation of the aggregated predictive flexibility offered by the electrical loads. Virtual battery (VB) models are often used to quantify the predictive flexibility in thermostatic loads (e.g. residential air-conditioners, electric water-heaters), which model the temporal evolution of a (virtual) energy state via a first order dynamics including self-dissipation rate, and power and energy capacities as parameters. Uncertainties and lack of information regarding end-usage and equipment models render deterministic VB models impractical. In this paper, we introduce the notion of stochastic VB models, and propose a \textit{variational autoencoder}-based deep learning algorithm to identify the probability distribution of the VB model parameters. Using available sensors and meters data, the proposed algorithm generates not only point estimates of the VB parameters, but also confidence intervals around those values. Effectiveness of the proposed frameworks is demonstrated on a collection of electric water-heater loads, whose operation is driven by uncertain water usage profiles.
\end{abstract}

\section{INTRODUCTION}
Advanced sensing, controls and communications infrastructure, especially at the medium-to-low voltage power distribution networks, have enabled the proliferation of connected, \textit{smart} appliances which are able to communicate with each other and/or a resource coordinator. In particular, Internet-of-Things (IoT) devices, such as smart thermostats and sensors, are capable of operating interactively and autonomously while remaining connected with other devices and/or the building automation system, and are often enabled by low cost cloud and computing platforms for local in-device data analytics and controls implementation \cite{manic2016intelligent}. It is becoming increasingly feasible to engage and coordinate these distribution side end-use resources to provide grid ancillary support \cite{Kirby:2002,callaway2009tapping,kundu2011modeling,kats2012buildings,perfumo2012load,Sinitsyn:2013,Mathieu:2013,zhang2013aggregated,Ma:13,hao2015aggregate,hughes2016identification,nandanoori2018prioritized}, thereby offering relatively faster, cleaner and cost-effective alternatives to more traditional measures. In order to deploy these resources for grid ancillary services, it is important for the operators to have access to predictive models of aggregated flexibility offered by these resources over some duration in the future. Several works in recent years have explored different aspects of flexibility based on the types of grid service considered, types of devices involved, as well as the methods of actuation. For example, short-term response of aggregated thermostatic load  to set-point control has been modeled in \cite{callaway2009tapping,kundu2011modeling,perfumo2012load,Sinitsyn:2013,Mathieu:2013,zhang2013aggregated}; method to model aggregated feasible set of active and reactive power consumption of flexible end-use resources have been discussed in \cite{kundu2018approximating,nazir2018inner,kundu2019scalable,singhal2019volt}; while virtual battery (VB)-based models have been used in \cite{hao2015aggregate,hughes2016identification} for predicting the active power and energy flexibility.

Thermostatic loads (e.g. residential air-conditioners, electric water-heaters) are a type of `energy-driven' loads for which the end-use quality of service depends on the energy consumption over a duration. These loads can leverage the thermal energy to offer certain flexibility in temporarily changing their power consumption without compromising on end-use service quality. As such, the battery-like models to represent the flexibility of thermostatic loads is acquiring momentum in the community. 
Most VB models in the existing literature assume a linear model to represent the temporal evolution of the virtual energy state driven by changes in the power consumption, with limits placed on the power consumption and energy state \cite{nandanoori2019identification,mathieu2015arbitraging,hao2015aggregate,hughes2016identification,chakraborty2018virtual}. Different methods have been proposed to calculate the parameters of a VB model. Analytical closed-form approaches (e.g. \cite{hao2015aggregate}) and optimization-based methods (e.g. \cite{hughes2016identification,nandanoori2019identification}) assume availability of accurate end-use device-specific detailed models which are often unknown in reality. In \cite{mathieu2015arbitraging}, authors used a system identification method that involves running open-loop experiments on the devices (such as turning them all `on' or `off' at the same time), which can be prohibitive from grid reliability point of view (especially, since such actions can trigger nonlinear modes \cite{kundu2014nonlinear}). A novel application of deep learning methods was proposed in \cite{chakraborty2018virtual} for identifying the VB parameters from historical closed-loop response data. A stacked \textit{autoencoder} model (see \cite{hinton2006reducing,bengio2007greedy,baldi2012autoencoders,larochelle2009exploring} for details) was used to mimic the dimensional reduction problem by extracting a representation of the virtual energy state at the encoding dimension. A convolution-based \textit{long-short-term-memory} network (see \cite{hochreiter1997long} for details) was used for calculating VB model parameters from temporal evolution of the energy state, while \textit{transfer learning} methods (see \cite{chen2015net2net,romero2014fitnets} for details) were used for fast re-training of the network in response to time-varying changes in the load population (due to changing availability of end-use appliances).

A key drawback of the above mentioned VB modeling efforts, however, is that the proposed VB models are deterministic and cannot capture the effect of end-use (and other) uncertainties on the available demand flexibility. There is a need to expand the modeling capabilities to reveal the underlying resource uncertainty and unpredictability. For example, in \cite{vrakopoulou2019chance} authors proposed a chance-constrained optimal power-flow formulation that allocates reserves across the network using uncertainty-aware predictive flexibility models of loads. The main contribution of this paper is the introduction of a stochastic VB modeling framework that is capable of representing the end-use uncertainties via extending the traditional \textit{point estimates} of the VB parameters to probability distributions and associated confidence intervals.


In this work, we propose a \textit{variational autoencoder} (VAE) based machine learning algorithm (see \cite{pu2016variational} for details) that is capable of - 1) modeling the impact of end-use uncertainties on the virtual energy state, 2) discovering the uncertainty propagation patterns in its temporal evolution, and 3) identifying the probability distributions associated with the parameters of the VB model. The rest of the paper is organized as follows: Section\,\ref{S:problem} introduces the VB model and the problem of estimating the parameters with confidence intervals; before going into the technical details of the proposed VAE based framework in Section\,\ref{S:solution}; numerical results are provided in Section\,\ref{S:results}; with the article being concluded in the Section\,\ref{S:conclusion}.

\section{PROBLEM DESCRIPTION}\label{S:problem}

A virtual battery (VB) is typically modeled as a first order dynamical system that captures the temporal evolution of the virtual energy state driven by the power input, with constraints specified on the power input and the energy state \cite{hao2015aggregate,hughes2016identification,chakraborty2018virtual}, as follows: 
\begin{subequations}\label{E:VB_model}
\begin{align}
\dot{x}(t) &= - a x(t) - u(t)\,,\quad x(0) = x_0  \\
C_1&\leq x(t) \leq C_2,\\
P^- &\leq u(t) \leq P^+,
\end{align}\end{subequations}
where $x(t)$ denotes the virtual state of charge, with $x_0$ being the initial state of charge; $a$ denotes the self-dissipation rate; $u(t)$ acts as an input to the VB, typically denoting the power consumption above a nominal (or, baseline) power profile; $C_1$ and $C_2$ denote the lower and upper energy limits, respectively; while $P^{-}$ and $P^+$ are, respectively, the lower and upper power limits. Overall, the vector 
\begin{align*}
    \Phi = [x_0,a, C_1, C_2, P^- ,P^+]
\end{align*} 
denotes the set of VB parameters. The VB model is used to estimate the capability of a collection of flexible thermostatic loads in tracking certain regulation signals. For example, the model \eqref{E:VB_model} can be used to predict what regulation signals, $r(t)$\,, can be tracked by the VB, i.e. $u(t)=r(t)$\,, and for how long, without violating the bounds on the energy state \cite{hughes2016identification,nandanoori2019identification}. The virtual energy state acts as a proxy for the thermal energy associated with the thermostatic load. The bounds on the energy state ensure that the end-user comfort constraints (e.g. temperature lying within specified limits) are being satisfied. 

The power and energy limits in a VB are typically time-varying \cite{hao2015aggregate,mathieu2015arbitraging}, due to their dependence on time-varying factors such as the outside air temperature. In such cases, the notions of \textit{sufficient} and \textit{necessary} VB models, as proposed in \cite{hao2015aggregate}, are useful which help generate \textit{static} abstractions of the time-varying VB models that fit various application needs. Consider, for example, a VB model with time-varying power and energy limits, given by $C_1(t),C_2(t),P^-(t)$ and $P^+(t)$, which are bounded over a duration of interest $\mathcal{T}$, i.e.:
\begin{align*}
    \forall t\in\mathcal{T}:~C_1(t)\in[\underline{C}_1,\overline{C}_1],\,C_2(t)\in[\underline{C}_2,\overline{C}_2],~&\overline{C}_1\!<\!\underline{C}_2\\
    P^-(t)\in[\underline{P}^-,\overline{P}^-],\,P^+(t)\in[\underline{P}^+,\overline{P}^+],~&\overline{P}^-\!<\!\underline{P}^+.
\end{align*}
It is possible to synthesize the smallest (\textit{sufficient}) and the largest (\textit{necessary}) \textit{static} VB abstractions as follows:
\begin{align*}
    \text{(smallest)}&\quad \underline{\Phi}=[x_0,a, \overline{C}_1, \underline{C}_2, \overline{P}^- ,\underline{P}^+]\\
    \text{(largest)}&\quad \overline{\Phi}=[x_0,a, \underline{C}_1, \overline{C}_2, \underline{P}^- ,\overline{P}^+]
\end{align*}
which can be interpreted as: any regulation signal successfully tracked by the smallest static VB model is also guaranteed to be tracked by the time-varying VB model (hence, \textit{sufficient}); while any regulation signal the largest static VB model fails to track cannot be successfully tracked by the time-varying VB model (hence, \textit{necessary}).

Regardless, in this paper, we will focus our attention to the modeling of end-use and other uncertainties in the \text{static} (but stochastic) VB model parameters, while leaving the issue of temporal variability for later work. Existing VB models and identification methods, e.g. the optimization-based approach \cite{hughes2016identification}, the closed-form approximations \cite{hao2015aggregate}, or the deep learning techniques \cite{chakraborty2018virtual}, assume a deterministic scenario which does not allow systematic representation of the uncertainties driven by unpredictable end-user behavior. 
%
%
In order to capture the end-use uncertainties, we would like to generate, for each VB parameter, its probability distribution and estimate the \textit{most likely} value as well as identify a confidence interval around the estimated value. Mathematically,
\begin{subequations}
\begin{align}
    \forall \phi\in\Phi:\quad&\text{find }\,\phi^*,\phi_\varepsilon^-,\phi_\varepsilon^+\\
    \text{such that, }& \phi^*\text{ is the mode of the distribution},\\
    &p(\phi^-\leq\phi\leq\phi^+)\geq 1-\varepsilon\,,\\
    \text{and }& \,\phi^-\leq\phi^*\leq\phi^+\,.
\end{align}    
\end{subequations}
where $0\!<\!\varepsilon\!<\!1$ is a small positive scalar chosen to specify the confidence interval. Note that such a modeling framework aligns well with the chance-constrained optimal power-flow formulation used in \cite{vrakopoulou2019chance}.

\section{VARIATIONAL AUTOENCODR FRAMEWORK}\label{S:solution}

\subsection{Overview: Probabilistic Encoder and Decoder}
Before we can say that our model is representative of our dataset, we need to make sure that for every data point $X$ in the dataset, there is at least one setting of the latent variables which causes the model to generate something very similar to $X$. Formally, say we have a vector of latent variables $z$ in a single dimensional space $Z$ (representative of the VB energy state in our example) which we can easily sample according to some probability density function $p(z)$ defined over $Z$. Then, say we have a family of deterministic functions $f(z;\theta)$, parameterized by a vector $\theta$ in some space $\Theta$,where $f:Z\times \Theta\rightarrow X$. $f$ is deterministic, but if $z$ is random and $\theta$ is fixed, then $f(z;\theta)$ is a random variable in the space $X$ . We wish to optimize $\theta$ such that we can sample $z$ from $p(z)$ and with high probability $f(z;\theta)$ will be similar to $X$ in our original data-set. Now we define the previous description mathematically, by aiming to maximize the probability of each $X$ in the original dataset under the entire generative process, according to
\begin{equation}
    p(X)=\int p(X|z;\theta)p(z)dz.\label{var1}
\end{equation}
In the variational autoencoder (VAE) proposed in this work, the choice of output distribution is considered to be Gaussian, i.e., $p(X|z;\theta)=\mathcal{N}(X|f(z;\theta),\sigma^2I)$. In other words, it has mean $f(z;\theta)$ and covariance of the product of the identity matrix $I$ and a hyperparameter $\sigma$.


\subsection{Probabilistic Moments}\label{S:prob}
At this point, it is important to find the analytic expression of respective (mean,standard deviation) of the VB state (single dimensional) representation, given the (mean,standard deviation) of the input space, as we have normalized input data using its mean and standard deviation. Let us consider $\mathbf{X}\mathtt{\sim}\mathcal{N}(\mu_{X},\Sigma_{X})$ is the distribution of the input data $\mathbf{X}$. Input data $\mathbf{X}$, passes through a network of the form (Affine, Affine, Affine, Relu), before transforming to VB state ($z$) in the encoding space. For calculation simplicity, we break down this series of transformation into the following two-stepped structure $\mathbf{X}\xrightarrow[]{(\text{Affine},\text{Affine})} \mathbf{Y}\xrightarrow[]{(\text{Affine},\text{Relu})}$ $z$, then we can write $\mathbf{Y}=\mathbf{W}_{2}\Big(\mathbf{W}_{1}\mathbf{X}+\mathbf{B}_{1}\Big)+\mathbf{B}_{2}$, where $\mathbf{Y}\in\mathbb{R}^{150}$, $\mathbf{W}_{2}\in\mathbb{R}^{150\times 200}$, $\mathbf{B}_{2}\in\mathbb{R}^{150\times 1}$,
$\mathbf{W}_{1}\in\mathbb{R}^{200\times 295}$,
and $\mathbf{B}_{1}\in\mathbb{R}^{200\times 1}$. After some algebraic manipulations, equivalent (mean-standard deviation) for $\mathbf{Y}$, corresponding to $(\mu_{X},\Sigma_{X})$, can be written as:
\begin{gather}
  \Sigma_{Y}=\Sigma_{X},\label{sigma}\\
  \mu_{Y}=\mathbf{W}_{2}\mathbf{W}_{1} \mu_{X}+(1-\Sigma_{X})\Big(\mathbf{W}_{2}\mathbf{B}_{1}+\mathbf{B}_{2}\Big).\label{mu}
\end{gather}
Now using (\ref{mu}), we have designed $\mathbf{B}_{2}$ to enforce a zero mean distribution at $\mathbf{Y}$, i.e., $\mathbf{Y}\mathtt{\sim}\mathcal{N}(0,\Sigma_{Y})$. $\mathbf{B}_{2}$ is designed as :
\begin{gather}
    \mathbf{B}_{2}=-\frac{\mathbf{W}_{2}\mathbf{W}_{1} \mu_{X}}{(1-\Sigma_{X})}-\mathbf{W}_{2}\mathbf{B}_{1},\label{B4Tl}
\end{gather}
to achieve zero mean distribution at $\mathbf{Y}$.

Let us consider $\mathbf{Y}\mathtt{\sim}\mathcal{N}(\mu_{Y}=0,\Sigma_{Y})$ is the output in the $\mathbf{Y}$ space, as discussed before where $\mu_{Y}=0$ and $\Sigma_{Y}$ is the known mean and standard deviation of vector $\mathbf{Y}$. After passing $\mathbf{Y}$, through a network of the form (Affine, ReLU, Affine), the functional form of the network output is $z=q(\mathbf{Y})=\mathbf{W}_{4}\max(\mathbf{W}_{3}\mathbf{Y}+\mathbf{B}_{3},\mathbf{0}_{50})+\mathbf{B}_{4}$, where $\max(.)$ is an element-wise operator. For our application $q:\mathbb{R}^{150}\rightarrow \mathbb{R}$, $\mathbf{W}_{3}\in\mathbb{R}^{50\times 150}$, $\mathbf{W}_{4}\in\mathbb{R}^{1\times 50}$, $\mathbf{B}_{3}\in\mathbb{R}^{50\times 1}$,
$\mathbf{B}_{4}\in\mathbb{R}$, and $\mathbf{0}_{50}$ is a $50$-dimensional vector of zeros. Now we will state two theorems to give analytic expression of first and second statistical moments of $q(\mathbf{Y})$, where $\mathbf{Y}\mathtt{\sim}\mathcal{N}(\mu_{Y}=0,\Sigma_{Y})$ (see \cite{bibi2018analytic} for proof of these theorems).
\begin{Theorem}
(First Moment) For any function $q(\mathbf{Y})$,
\begin{gather}
    \mathbb{E}[q_{i}(\mathbf{Y})]=\sum_{j=1}^{50} \mathbf{W}_{4}(i,j)\bigg(\frac{1}{2}\mu_{j}-\frac{1}{2}\mu_{j}\textit{erf }\Big(\frac{-\mu_{j}}{\sqrt(2)\sigma_{j}}\Big)\nonumber\\+\frac{1}{\sqrt{2\pi}}\sigma_{j}\exp \Big(\frac{-\mu_{j}^{2}}{2\sigma_{j}^{2}}\Big)\bigg)+\mathbf{B}_{4}(i),\label{first_moment}
\end{gather}
where $\mu_{j}\triangleq \mathbf{W}_{3}\mu_{Y}+\mathbf{B}_{3}$, $\sigma_{j}\triangleq \bar{\Sigma} (j,j)$, $\bar{\Sigma}\triangleq  \mathbf{W}_{3}\Sigma_{Y}\mathbf{W}_{3}^{T}$ and $\textit{erf }(x)\triangleq\frac{2}{\sqrt{\pi}}\int_{0}^{x} e^{-t^{2}}dt$.
\end{Theorem}
\begin{Theorem}
(Second Moment) For any function $q(\mathbf{Y})$ where $\mathbf{Y}\mathtt{\sim}\mathcal{N}(0,\Sigma_{Y})$ and $\mathbf{B}_{3}=\mathbf{0}_{50}$, we get
\begin{gather}
    \mathbb{E}[q_{i}^{2}(\mathbf{Y})]=2\sum_{j_{1}=1}^{50}\sum_{j_{2}=1}^{j_{1}-1} \mathbf{W}_{4}(i,j_{1})\mathbf{W}_{4}(i,j_{2})\nonumber\\\bigg(\frac{\sigma_{j_{1},j_{2}}}{2\pi}\sin^{-1}{\Big(\frac{\sigma_{j_{1},j_{2}}}{\sigma_{j_{1}}\sigma_{j_{2}}}\Big)}+\frac{\sigma_{j_{1}}\sigma_{j_{2}}}{2\pi}\sqrt{1-\frac{\sigma^{2}_{j_{1},j_{2}}}{\sigma^{2}_{j_{1}}\sigma^{2}_{j_{2}}}}\nonumber\\+\frac{\sigma_{j_{1},j_{2}}}{4}\bigg)+\frac{1}{2}\sum_{r=1}^{50}\mathbf{W}_{4}(i,r)^{2}\sigma^{2}_{r}+\mathbf{B}_{4}(i).\label{second_moment}
\end{gather}
\end{Theorem}

The mean $\mu_{z}$ and standard deviation $\Sigma_{z}$ of single dimensional VB state $Z$ can be calculated using (\ref{first_moment}) and (\ref{second_moment}) as, $\mu_{z}\triangleq \mathbb{E}[q_{i}(\mathbf{Y})]$ and $\Sigma_{z}\triangleq \sqrt{\mathbb{E}[q_{i}^{2}(\mathbf{Y})]-\Big(\mathbb{E}[q_{i}(\mathbf{Y})]\Big)^{2}}$\,. In the following, we describe the framework for handling the end-use uncertainties.

\subsection{Proposed VAE}\label{S:VAE}
The key idea behind training our proposed VAE is to attempt to sample $z$, which is likely to produce a data point similar to $X$, and simultaneously compute $p(X)$, just for the sample $z$. We now define a function $q(z|X)$ which can take a value of $X$ and give us a distribution over $z$ values that are likely to produce similar data points as in $X$. We define Kullback-Leibler (KL) divergence between $p(z|X)$ and $q(z)$, for some arbitrary $q$, as 
\begin{equation}
    \mathcal{D}[q(z)||p(z|X)=E_{z\sim q}[\mathrm{log\,}q(z)-\mathrm{log\,}p(z|X)].\label{var2},
\end{equation}
and rewrite it by applying the Bayes' rule to $p(z|X)$ 
\begin{gather}
    \mathcal{D}[q(z)||p(z|X)=E_{z\sim q}[\mathrm{log\,}q(z)-\mathrm{log\,}p(X|z)\nonumber\\-\mathrm{log\,}p(z)]+\mathrm{log\,}p(X).\label{var3}
\end{gather}
Rearranging terms and re-applying KL-divergence, we get 
\begin{gather}
    \mathrm{log\,}p(X)-\mathcal{D}[q(z)||p(z|X)]=E_{z\sim q}[\mathrm{log\,}p(X|z)]\nonumber\\-\mathcal{D}[q(z)||p(z)].\label{var4}
\end{gather}
Since we want to infer $p(X)$, we construct $q$ which does not depend on $X$, and in particular makes $\mathcal{D}[q(z)||p(z|X)]$ small, i.e.,
\begin{gather}
   \mathrm{log\,}p(X)-\mathcal{D}[q(z)||p(z|X)]=E_{z\sim q}[\mathrm{log\,}p(X|z)]\nonumber\\-\mathcal{D}[q(z|X)||p(z)].\label{var5}
\end{gather}
We want to maximize the left hand side of Equation \ref{var5}, while we also want to optimize the right hand side of Equation \ref{var5} using stochastic-gradient descent, given right choice of $q$. Moreover, the right hand side of Equation \ref{var5} behaves similar to an autoencoder, where $q$ encodes $X$ into $z$ and $P$ decodes it to reconstruct $X$.



In order to perform stochastic-gradient descent on the right side of Equation \ref{var5}, we choose $q(z|X)=\mathcal{N}(z|\mu(X;\theta),\Sigma(X;\theta))$, where $\mu$ and $\Sigma$ are deterministic functions with parameters $\theta$, that can be learned from data. In our framework $\Sigma$ is constrained to be a diagonal matrix. Because of the choice of $q$, the last term of the right hand side becomes
\begin{gather}
    \mathcal{D}[q(z|X)||p(z)]=\frac{1}{2}(\mathrm{tr}(\Sigma(X))+(\mu(X))^{T}(\mu(X)-k\nonumber\\-\mathrm{log\,}(\mathrm{det}(\Sigma(X))))),\label{var6}
\end{gather}
where $k$ is the dimensionality of the distribution. Finally our objective function for optimization associated with training of VAE can be written as
\begin{align}
    E_{X\sim D}[\mathrm{log\,}p(X)-\mathcal{D}[q(z|X)||p(z|X)]]=\nonumber\\E_{X\sim D}[E_{z\sim q}[\log p(X|z)]
    -\mathcal{D}[q(z|X)||p(z)]].\label{var7}
\end{align}

Now we evaluate performance of proposed VAE and consequently identify VB parameter ($\phi$, as introduced in Section \ref{S:problem}) using the Probabilistic Moments calculation introduced in Section \ref{S:prob}.

\section{NUMERICAL EXAMPLE}\label{S:results}
\subsection{Dataset Description and Dataset Splitting}
\label{sec3.1}
\begin{figure}
\centering
\includegraphics[width=\linewidth]{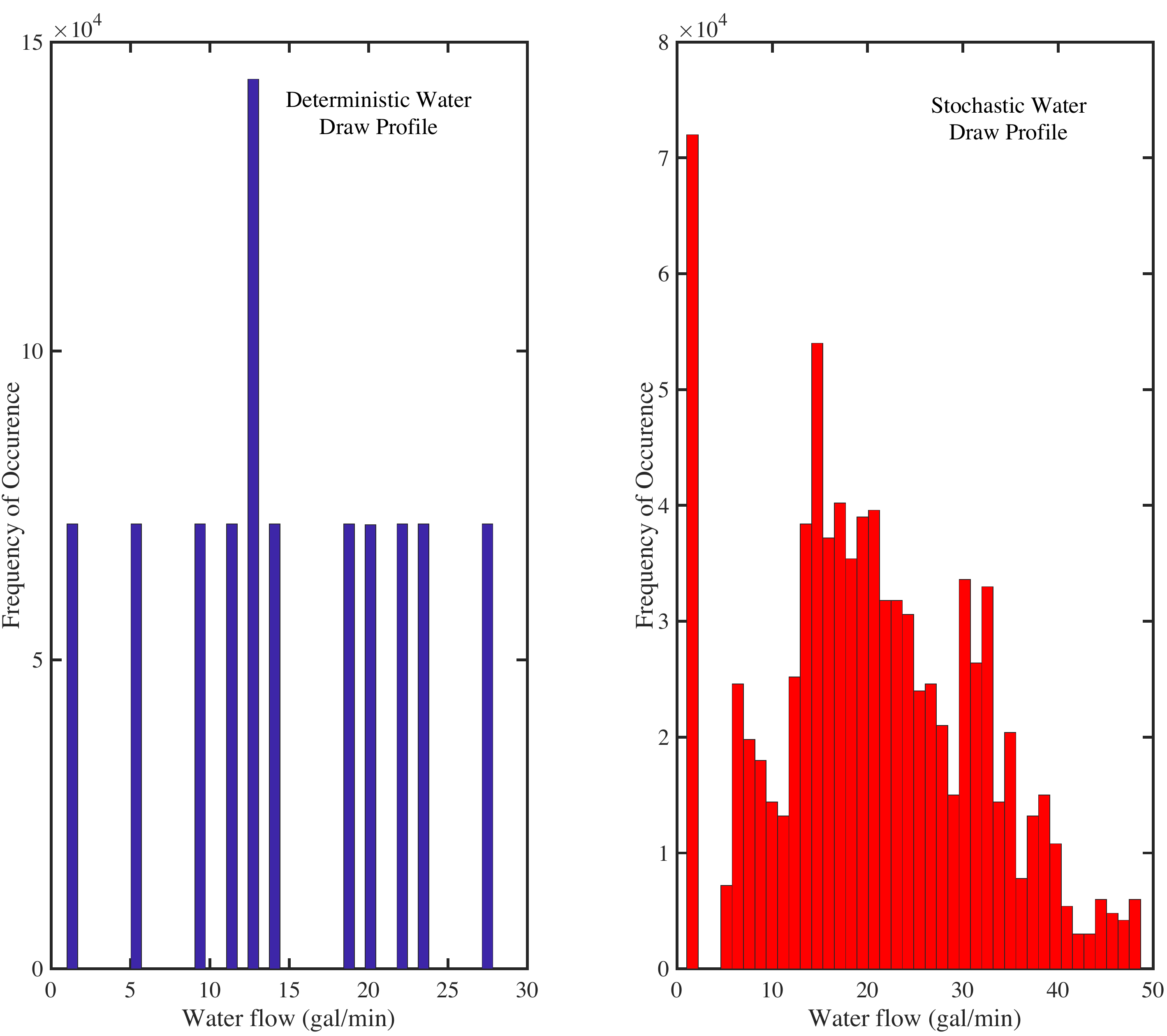}
\caption{Stochastic Water draw profile used for EWH, used for training the proposed VAE. An example water profile is also shown without the presence of any uncertainty.}
\label{fig:water}       
\end{figure}

For numerical illustration of the proposed VAE-based framework, we consider an ensemble of 150 electric water heater (EWH) devices. The regulation (tracking) signals from PJM \cite{PJM_signals} are considered and scaled appropriately to match the ensemble of EWHs. The EWHs in the ensemble change their operational state (ON/OFF) in order to track a regulation signal. However, during the tracking process, the switching actions on the EWHs are to be performed so as to not violate the local end-use specified temperature constraints. In this paper, we implement the switching strategy as the solution of an optimization problem proposed in \cite{chakraborty2018virtual}. In this process, we generated and collected the time-series data of the temperature of each of the 150 EWHs for a 2 hours duration, at 1 second time resolution, for $200$ distinct regulation signals. If the ensemble fails to track a regulation signal, then the time-series data is considered up to the point where tracking fails. The parametric uncertainty is considered in the water draw profile, and Fig. \ref{fig:water} shows an example water draw profile, with and without uncertainty.
The power limits of the ensemble are computed through a one-sided binary search algorithm as described in 
\cite{nandanoori2019identification}. For some regulation signals the ensemble violates the power limits $P^-$ and $P^+$ before the $2$ hour running time and only the temperature of each EWH is considered, until the time when the ensemble satisfies the power limit. Finally, for making a suitable dataset for applying the proposed VAE, we stack the temperature of each EWH device, followed by temperature set points for each device, by column, and then stack the data points for each regulation signal by row. For the selected ensemble, this stacking results in a dataset of dimension $\mathbb{R}^{1440199\times 295}$. 
We have used a 10-fold cross validation for training and validation of our proposed VAE. We have kept the testing set separated as an indicator of generalized performance. For the given dataset (Section \ref{sec3.1}) $30\%$ of the dataset is separated and kept as a test dataset. The remaining $70\%$ of the dataset has been used in the random cross validation, for both training and validation of the proposed VAE.

\subsection{Results}
\begin{figure}[!ht]
     \centering
	\includegraphics[width=0.8\columnwidth]{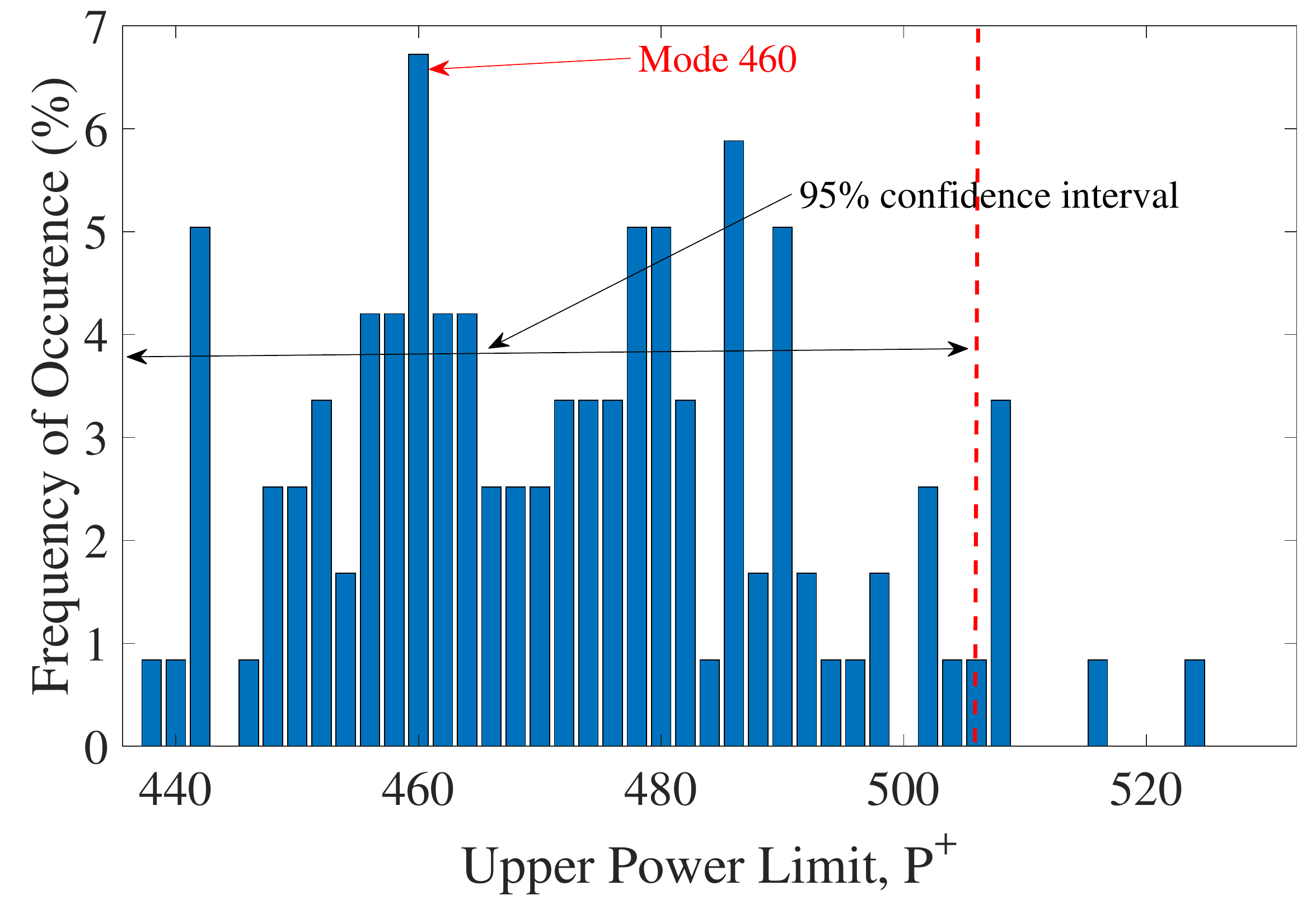}      \caption{Distribution of $P^{+}$ in kW, with mode at $460$ kW and $95\%$ confidence interval}
      \label{Fig1}
\end{figure}
\begin{figure}[!ht]
     \centering
     \includegraphics[width=0.8\columnwidth]{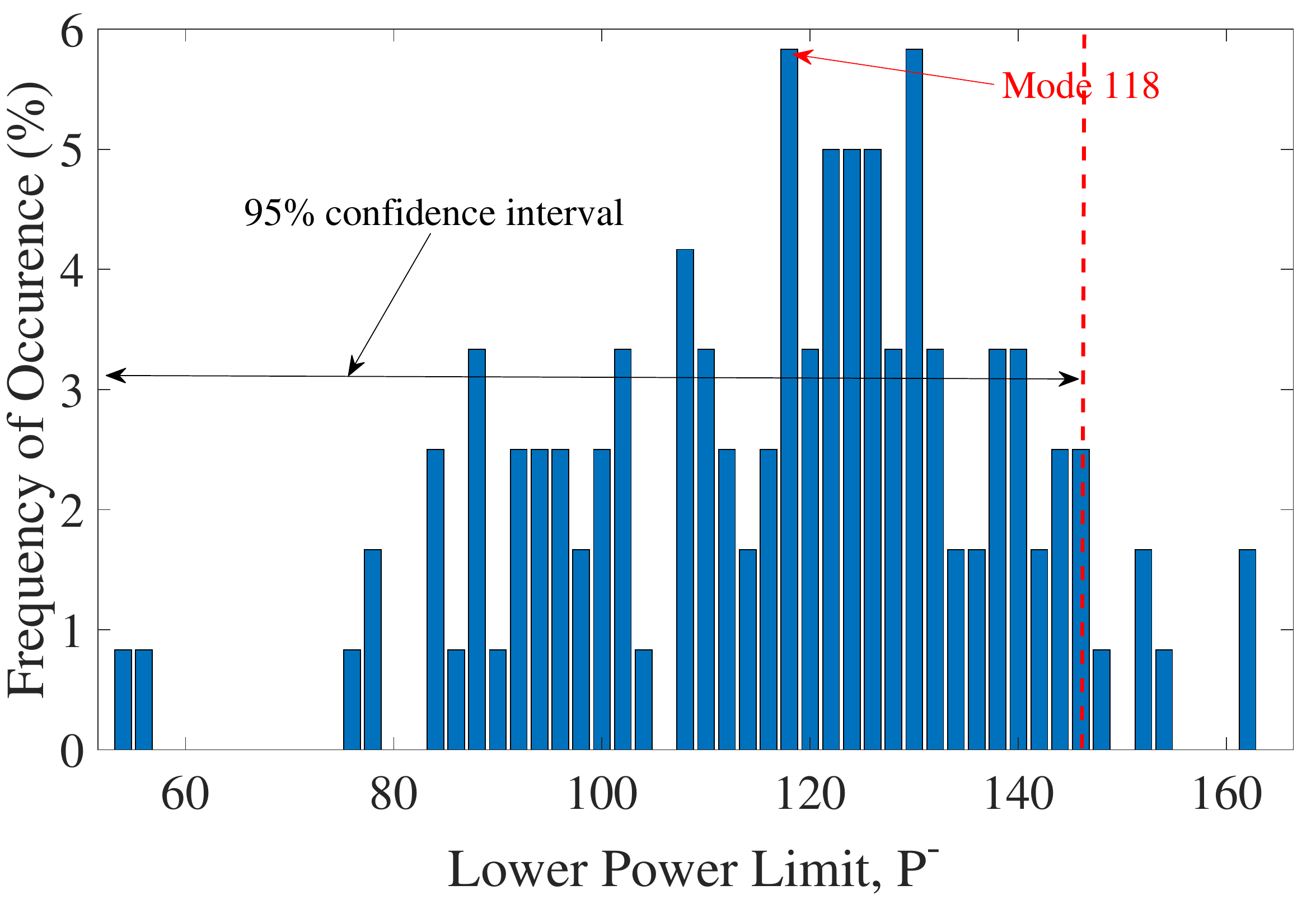}      \caption{Distribution of $P^{-}$ in kW, with mode at $118$ kW and $95\%$ confidence interval}
      \label{Fig2}
\end{figure}
\begin{figure}[!ht]
     \centering
	\includegraphics[width=0.95\columnwidth]{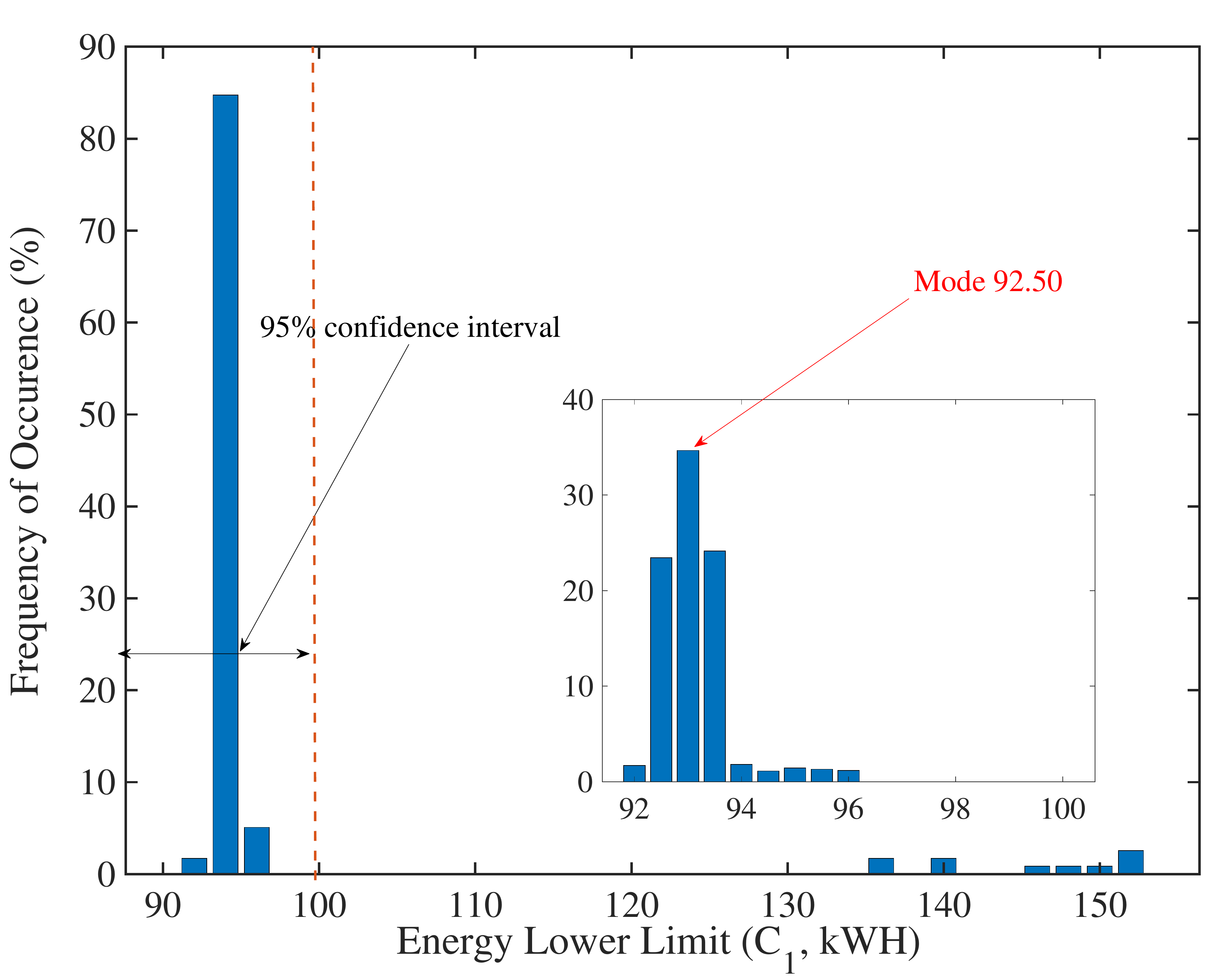}      \caption{Distribution of lower energy limit $C_{1}$, with mode at $92.5$ kWh and $95\%$ confidence interval}
      \label{Fig3}
\end{figure}
\begin{figure}[!ht]
     \centering
	\includegraphics[width=0.95\columnwidth]{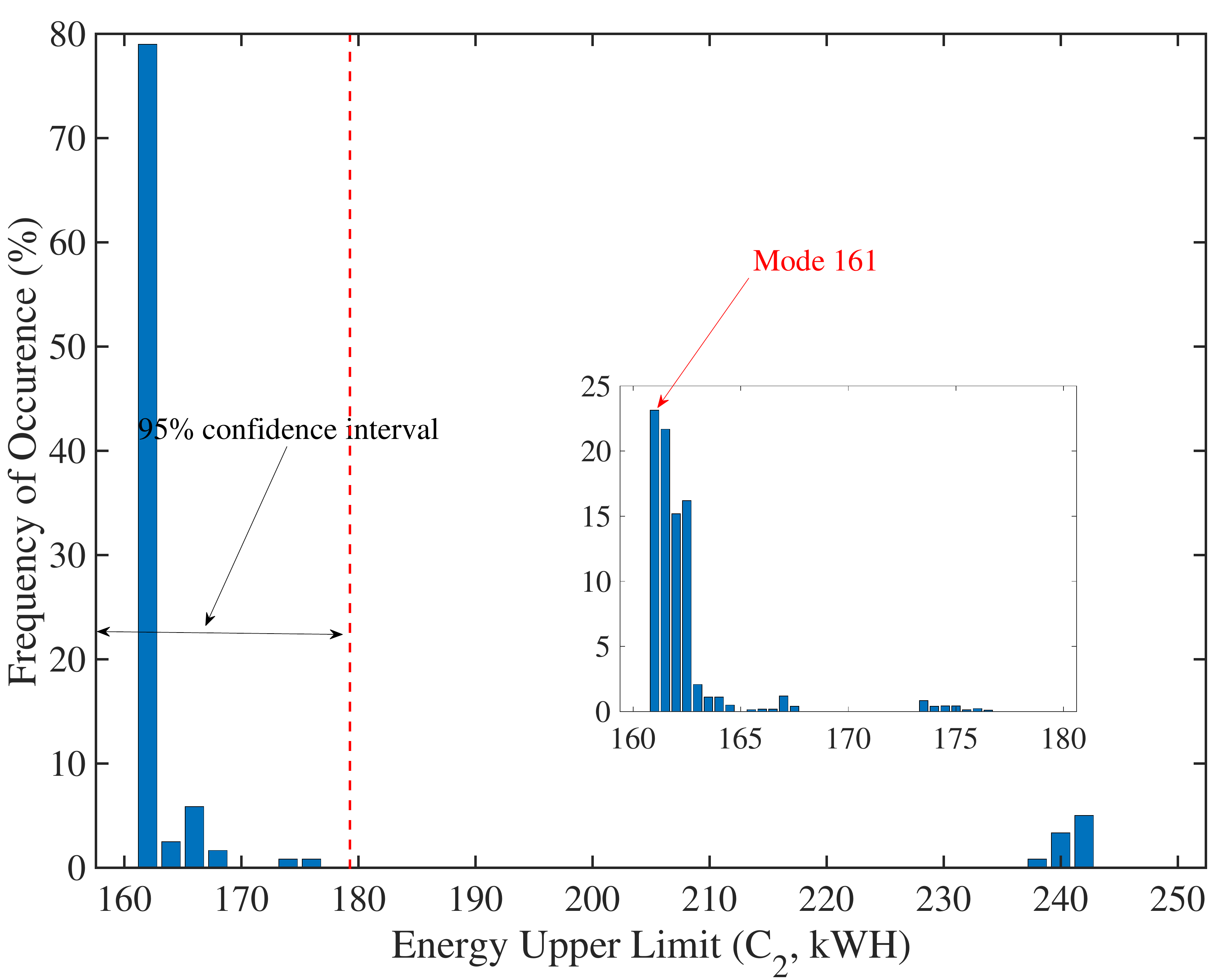}      \caption{Distribution of upper energy limit $C_{2}$, with mode at $161$ kWh and $95\%$ confidence interval}
      \label{Fig4}
\end{figure}
\begin{figure}[!ht]
     \centering
	\includegraphics[width=0.95\columnwidth]{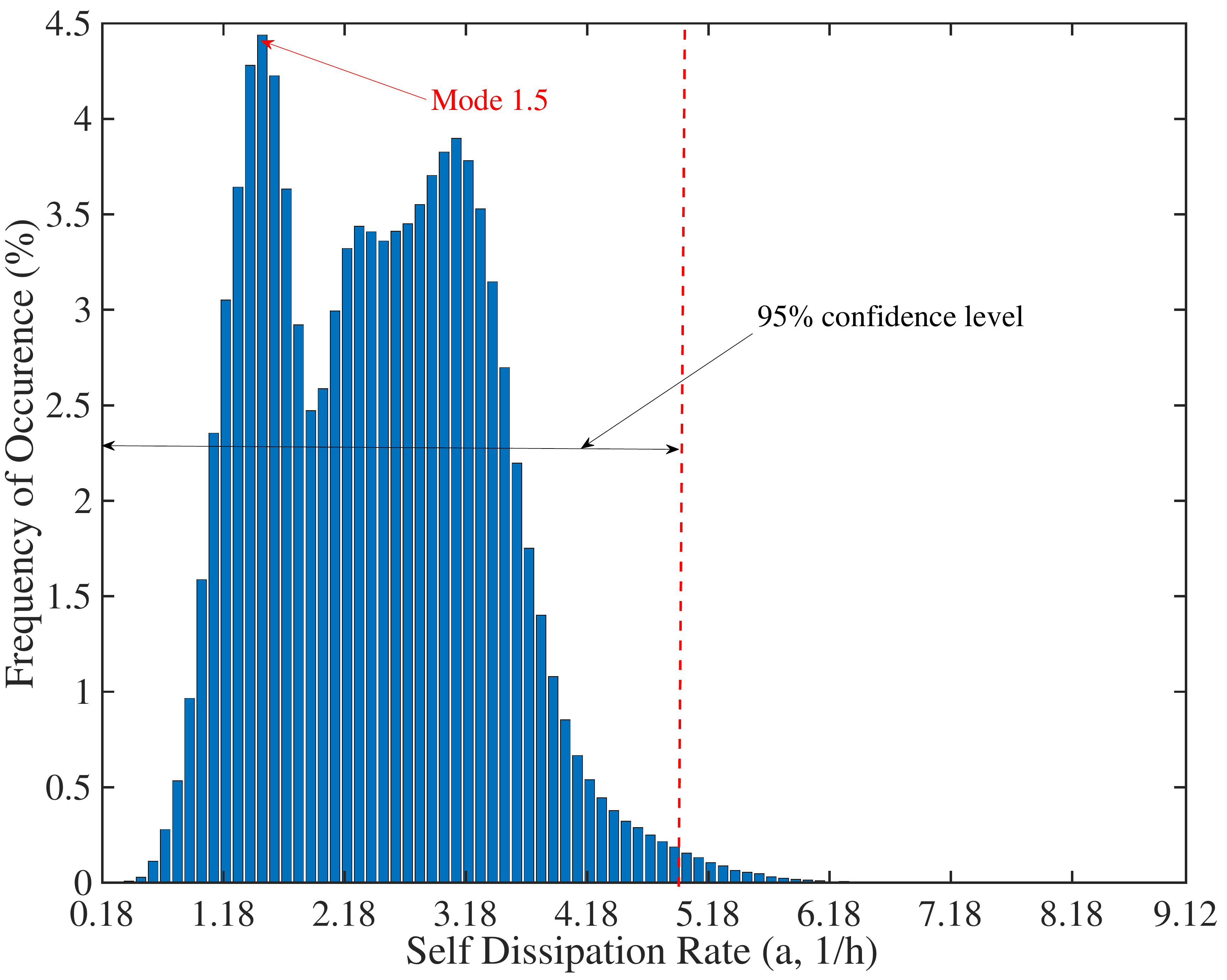}      \caption{Distribution of self dissipation rate $a$, with mode at $1.5$ 1/h and $95\%$ confidence interval}
      \label{Fig5}
\end{figure}
In Fig. \ref{Fig2}-\ref{Fig5}, we have plotted the distribution we found from the latent space of the trained VAE, for different identifiable VB parameters, $\phi$. These plots also indicates the mode value of each of these variables, along with the $95\%$ confidence interval values. We want to emphasis here that the parametric uncertainty of EWH devices as shown in Fig. \ref{Fig1} is not fitted in the VAE. The latent representation of the VAE, along with the probabilistic moment calculation is able to identify the distribution of different parameters in $\phi$. 

\begin{figure}[!ht]
     \centering
	\includegraphics[width=\columnwidth]{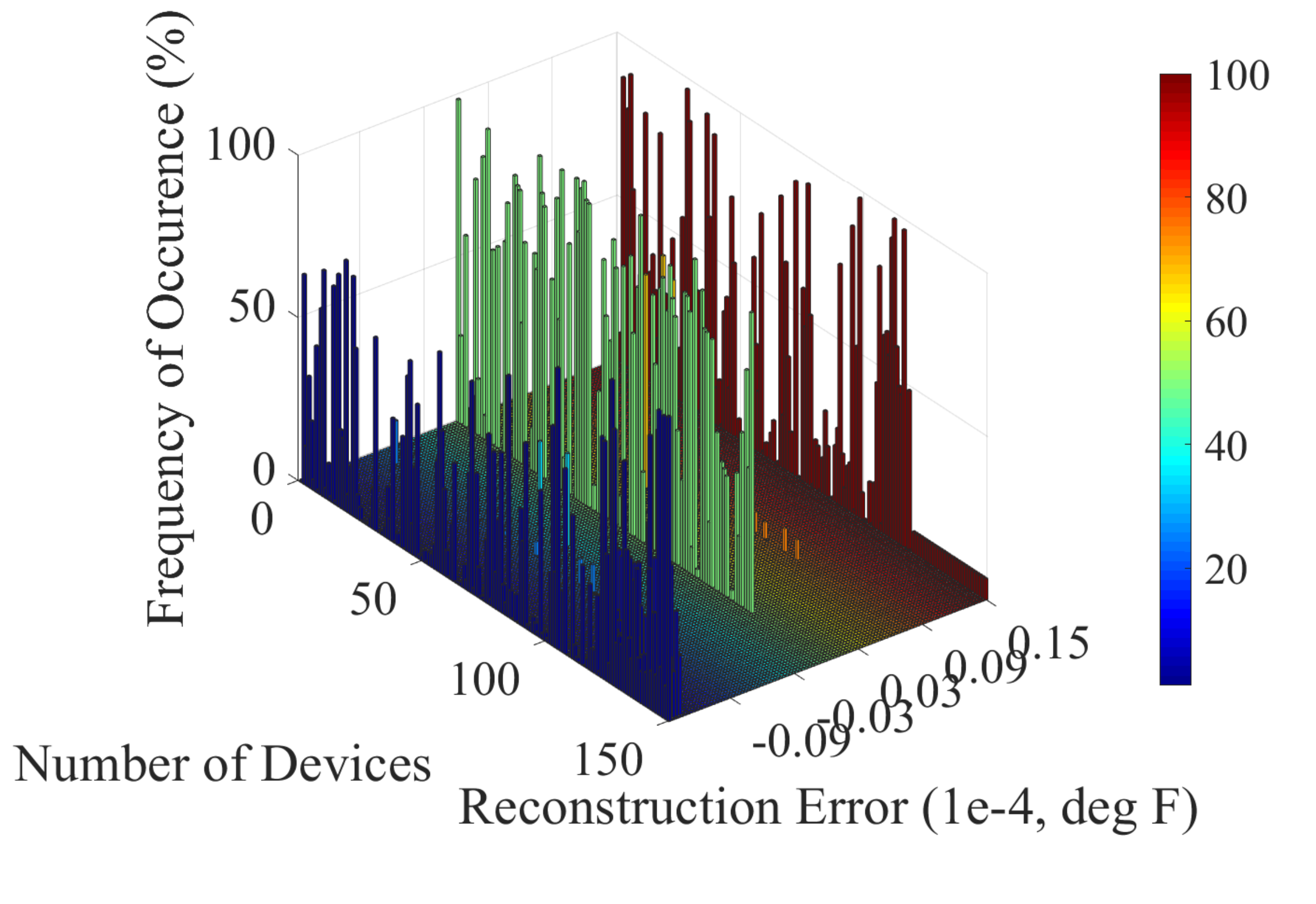}      \caption{Reconstruction Error in $^{\circ}$F for different WH devices and their associated distribution}
      \label{Fig6}
\end{figure}

The reconstruction loss is plotted in Fig. \ref{Fig6} for 150 different EWH devices, and as we can notice the maximum reconstruction loss is in the order of 0.15$^{\circ}$\,F, which is less than $0.10\%$ of reconstruction error (maximum reaching temperature for the EWH devices is 120$^{\circ}$\,F). Also, as the colorbar indices the frequency of occurrence of different reconstruction error value for different EWH devices, we can easily notice that majority of the reconstruction losses are very close to zero for all the EWH devices.

\begin{figure}[!ht]
     \centering
	\includegraphics[width=\columnwidth]{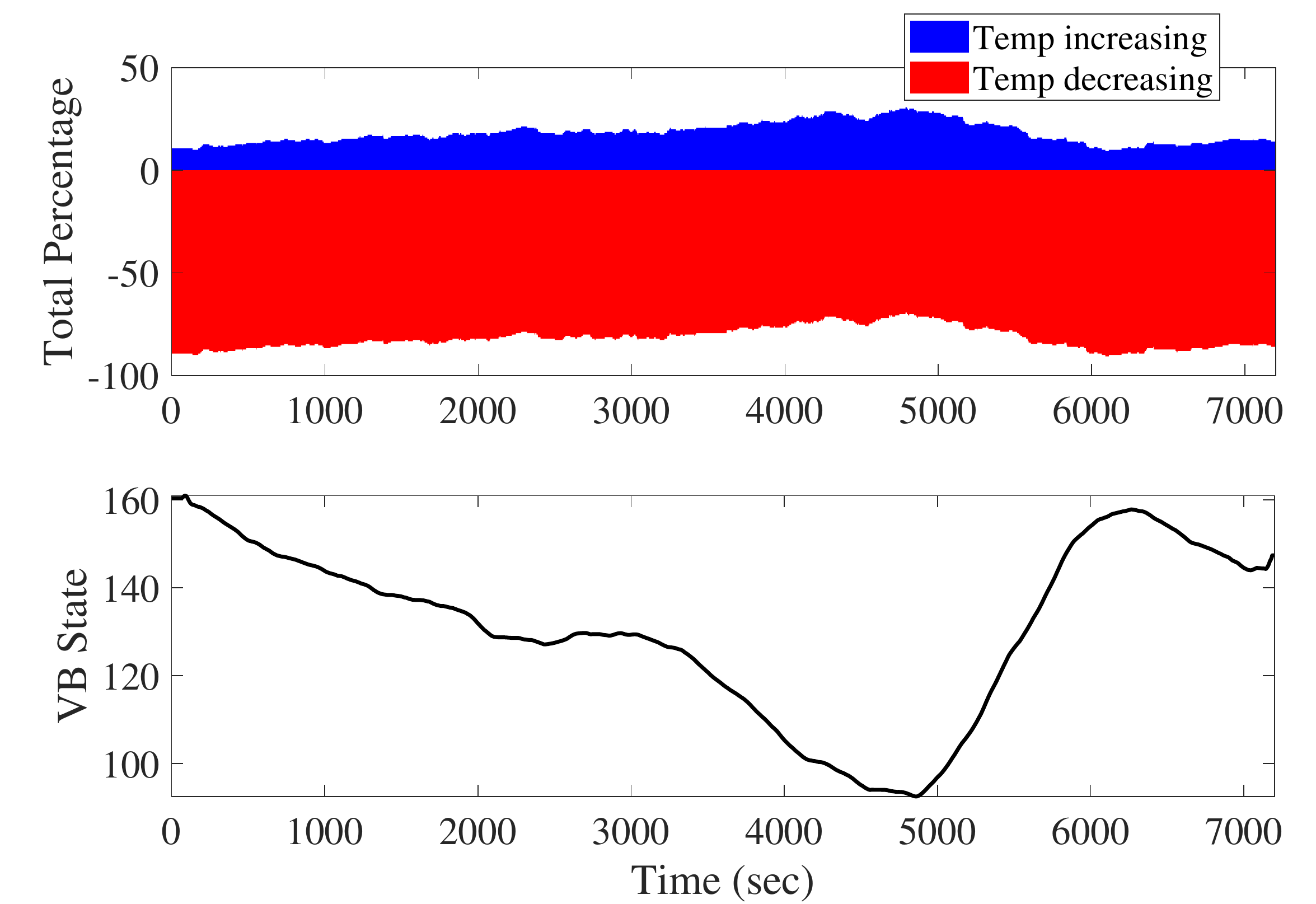}      \caption{Percentage of devices having increments and decrements in the temperature profile, along with the evolution of VB state with time.}
      \label{Fig7}
\end{figure}

Finally, in Fig. \ref{Fig7} we have tried to draw an intuitive connection between the device temperature changes and the change in VB state. As showed in Fig. \ref{Fig7}, the red and blue color indicates number of devices with decreasing and increasing temperature, respectively, which when added gives a similar trend of profile as of the identified VB state.

\section{CONCLUSION}\label{S:conclusion}
\vspace{-5pt}
Variational autoencoder (VAE) gained popularity due to its inherent property of identifying distribution in its latent space. We have utilized VAE in the context of identifying the virtual battery (VB) state for an ensemble of thermostatic loads, subjected to parametric uncertainty. We propose an extension of the deterministic VB model available in the literature to a stochastic one in order to better represent the end-use uncertainties that govern the available flexibility in thermostatic loads. Using a novel application of VAE, we illustrate how the end-use uncertainties can be captured by generating probability distributions of the VB parameters. Finally we have evaluated the performance of our framework in the context of an ensemble of $150$ electric water heater devices subjected to uncertainties in the water-draw profile. Although we have evaluated our proposed VAE based framework for parametric type uncertainties, it can be extended for other types such as modeling uncertainty.

\bibliographystyle{IEEEtran}
\bibliography{MyReferences}

\end{document}